\documentclass[copyright,creativecommons]{eptcs}
\providecommand{\event}{RTRTS 2010} 
\usepackage{breakurl}

\usepackage{latexsym} 
\usepackage{amsmath}
\usepackage{amssymb}
\usepackage{amsfonts}

\usepackage{stmaryrd}
\usepackage{mathrsfs}

\usepackage{url}
\usepackage{alltt}

\usepackage{graphicx}
\usepackage{color}

\usepackage{shortvrb}  
\usepackage{dsfont}    

\usepackage{epsfig}

\usepackage{longtable}
\usepackage{booktabs} 
\newcommand{\ra}[1]{\renewcommand{\arraystretch}{#1}}

\usepackage{subfig}

\newcommand{\attone}{\mbox{$att_1$}}
\newcommand{\attn}{\mbox{$att_n$}}
\newcommand{\sone}{\mbox{$s_1$}}
\newcommand{\sn}{\mbox{$s_n$}}

\newcommand{\ignore}[1]{}

\newcommand{\hide}[1]{}

\def\myttsize{\small}

\begin{document}

\title{A Rewriting-Logic-Based Technique for \\ Modeling   Thermal Systems}
\author{Muhammad Fadlisyah 
\institute{University of Oslo}
\and Erika \'Abrah\'am 
\institute{RWTH Aachen}
\and Daniela Lepri 
\institute{University of Oslo}
\and Peter Csaba {\"O}lveczky
\institute{University of Oslo}
}

\def\titlerunning{A Rewriting-Logic-Based Technique for Modeling  Thermal Systems}

\def\authorrunning{M. Fadlisyah et al.}

\maketitle

\begin{abstract}
This paper  presents a  rewriting-logic-based modeling and
analysis technique for physical systems, with focus on
thermal systems. 
The  contributions of this paper can be summarized as follows:
(i) providing a framework for modeling and executing physical systems, where 
 both  the physical components and  their physical interactions are
 treated as first-class citizens;       
(ii) showing how heat transfer problems in thermal systems can be
modeled in Real-Time Maude;      
(iii) giving the implementation in Real-Time Maude of a basic
numerical technique for  executing continuous behaviors in
object-oriented hybrid systems; and  
(iv) illustrating these techniques with a set of incremental case
studies using realistic physical parameters, with examples of 
simulation and model checking analyses.

\end{abstract}

\MakeShortVerb{\@}

\section{Introduction}\label{sec:introduction}

The rewriting-logic-based \emph{Real-Time Maude} tool~\cite{journ-rtm}
has proved to be very useful for formally modeling and analyzing a
wide range of advanced real-time systems (see,
e.g.,~\cite{mike-wsn,norm-paper,fase06,aer-journ,ogdc-tcs}) that are
beyond the scope of timed-automaton-based tools. One important
question to investigate is to what degree -- and \emph{how} --
Real-Time Maude can be successfully applied to formally model and
analyze \emph{hybrid} systems.

As part of this investigation, this paper presents a
rewriting-logic-based modeling and analysis technique for
\textit{physical systems}, which 
consist of a set of \emph{components} that behave and interact
according to the laws of physics. On the one hand, the physical
quantities of the components \textit{continuously} evolve in time. On
the other hand, the components may also have \textit{discrete} state
parts, thus exhibiting \textit{hybrid} behaviors. 

We propose a technique to generate \emph{executable
  models} of physical systems in Real-Time Maude. Our modeling
technique is based on an adaptation of the \textit{effort} and
\textit{flow} approach described in \cite{wellstead}, explicitly
modeling the transfer of power between the components.  For the
continuous behavior of physical systems, which is described by
differential equations, the execution is based on the Euler method
(see, e.g.,~\cite{hoffman}) giving approximate solutions to ordinary
differential equations. This numerical method operates with a discrete
sampling strategy over a continuous time domain.


This paper focuses on \textit{thermal systems}, which are physical
systems that deal with  heat transfer and temperature change
problems. 
Thermal systems appear in many important computer-controlled
(or computer-controllable) 
applications,  such as  house heating systems, human body
thermoregulatory systems, and nuclear power plants.  
%
We describe the use of our modeling and execution technique with a 
sequence of increasingly complex thermal systems.  
We first model simple thermal systems, such as ``a cup of coffee in a
room''; then we add a coffee heater to the system; and, finally,   we add
an automatic coffee heater as a controlling system that manages the
coffee temperature.

The main contributions of this paper can be summarized as follows:
\begin{itemize}
\item We develop a framework for \emph{modeling} general
  \emph{physical systems} in Real-Time Maude.
\item This framework is used as the basis to define a technique 
  to model heat transfer in \emph{thermal systems}.
  We provide some basic classes, equations, and rules that can be used
  or extended to build models of thermal systems.
\item The implementation of a basic numerical technique
 to execute continuous behavior of
  object-oriented real-time systems, which allows the analysis of
  their hybrid behavior.
\item Incremental \emph{case studies}, using realistic physical
  parameters, with examples of simulation and reachability analysis.
\end{itemize}

\par
We give an overview of Real-Time Maude in Section~\ref{sec:rtm}.
Section~\ref{sec:thermalsys} gives a brief introduction to~thermal
systems.  Section~\ref{sec:modeling_framework} presents a high-level
overview of modeling physical and thermal systems with the effort and
flow approach, and approximating their behavior.
Section~\ref{sec:modeling_thermalsys} describes our modeling and
execution framework for thermal systems using Real-Time Maude.  Several
case studies are given in Section~\ref{sec:case-studies}.
Section~\ref{sec:related-work} discusses related work, and 
Section~\ref{sec:concl} mentions future work and summarizes the~paper.

\section{Real-Time Maude}
\label{sec:rtm}

In this section we briefly introduce Real-Time Maude~\cite{journ-rtm}, a rewriting-logic-based tool
supporting the formal  modeling and analysis of real-time systems.
A Real-Time Maude \emph{timed module} specifies a
\emph{real-time rewrite theory}   $(\Sigma, E, \mathit{IR},
\mathit{TR})$, where:  

\begin{itemize}
\item $(\Sigma, E)$ is a \emph{membership equational
    logic}~\cite{maude-book} theory with $\Sigma$ a
  signature\footnote{i.e., $\Sigma$ is a set of declarations of
    \emph{sorts}, \emph{subsorts}, and \emph{function symbols}} and
  $E$ a set of {\em confluent and terminating conditional equations}.
  $(\Sigma, E)$ specifies the system's state space as an algebraic
  data type, and must contain a specification of a sort @Time@
  modeling the (discrete or dense) time domain.

\item $\mathit{IR}$ is a set of (possibly conditional) \emph{labeled
    instantaneous rewrite rules} of the form

  {  \small
\begin{alltt}
       rl [\(l\)] : \(t\) => \(t'\) 
      crl [\(l\)] : \(t\) => \(t'\) if \(cond\)
\end{alltt}}  

\noindent   specifying the system's
  \emph{instantaneous} (i.e., zero-time) one-step transitions from an
  instance of $t$ to the corresponding instance of $t'$, where $l$ is
  a \emph{label}.  Conditional instantaneous rules apply only  if their
  conditions hold. 
  The rules are applied \emph{modulo} the equations~$E$.\footnote{$E$
    is a union $E'\cup A$, where $A$ is a set of equational axioms
    such as associativity, commutativity, and identity, so that
    deduction is performed \emph{modulo} $A$.  Operationally, a term
    is reduced to its $E'$-normal form modulo $A$ before any rewrite
    rule is applied.}

\item $\mathit{TR}$ is a set of (possibly conditional) \emph{tick
    (rewrite) rules}, written with syntax

  {\small
\begin{alltt}
       rl [\(l\)] : \texttt{\char123}\(t\)\texttt{\char125} => \texttt{\char123}\(t'\)\texttt{\char125} in time \(\tau\) \qquad
      crl [\(l\)] : \texttt{\char123}\(t\)\texttt{\char125} => \texttt{\char123}\(t'\)\texttt{\char125} in time \(\tau\) if \(cond\)
\end{alltt}}

\noindent that model time elapse.  @{_}@ is a 
built-in
 constructor of  sort \texttt{GlobalSystem}, and
$\tau$ is a term of sort @Time@ that denotes the \emph{duration}
of the rewrite.
\end{itemize}
The initial state must be a ground term of sort @GlobalSystem@ and 
 must be reducible to a term of
the form @{@$t$@}@ using the equations in the specifications. The form
of
the tick rule then ensures that time advances uniformly in all parts
of the system. 

The Real-Time Maude syntax  is fairly intuitive. For example, 
 a function symbol $f$ is 
  declared with the syntax \texttt{op }$f$ @:@ $s_1$ \ldots $s_n$
 @->@ $s$, where
$s_1\:\ldots\:s_n$ are the sorts of  its arguments,  and $s$ 
 is its (value) \emph{sort}. Equations are written
with syntax @eq@ $t$ @=@ $t'$, and @ceq@ $t$ @=@ $t'$ @if@ \emph{cond}
for conditional equations. The mathematical variables in such statements
are declared with the keywords {\tt var} and {\tt vars}.
We refer to~\cite{maude-book} for more details  on the syntax of
 Real-Time Maude. 


In object-oriented Real-Time Maude modules, a \emph{class} declaration

\myttsize
\begin{alltt}
  class \(C\) | \(\attone\) : \(\sone\), \dots , \(\attn\) : \(\sn\) .
\end{alltt}
\normalsize 

\noindent declares a class $C$ with attributes $att_1$ to $att_n$ of
sorts $s_1$ 
to $s_n$. An {\em object\/} of class $C$ in a  given state is
represented as a term
$@<@\: O : C \mid att_1: val_1, ... , att_n: val_n\:@>@$
of sort @Object@, where $O$, of sort @Oid@,  is the
object's
\emph{identifier}, and where $val_1$ to 
$val_n$ are the current values of the attributes $att_1$ to 
$att_n$.
 In a concurrent object-oriented
system, the  
 state
 is a term of 
the sort @Configuration@. It  has 
the structure of a  \emph{multiset} made up of objects and 
\emph{messages} (that are terms of sort \texttt{Msg}).
Multiset union for configurations is denoted by a juxtaposition
operator (empty
syntax) that is declared associative and commutative, so that rewriting is 
\emph{multiset
rewriting} supported directly in Real-Time Maude.

The dynamic behavior of concurrent
object systems is axiomatized by specifying its concurrent
transition patterns by rewrite rules. For example, 
the rule

{\small
\begin{alltt}
  rl [l] :  < O : C | a1 : x, a2 : y, a3 : z >  
            < O' : C | a1 : w, a2 : 0, a3 : v >
           =>
            < O : C | a1 : x + w, a2 : y, a3 : z >  
            < O' : C | a1 : w, a2 : x, a3 : v > 
\end{alltt}
}

\noindent  defines a family of transitions 
 where two objects of class @C@
synchronize  to update their attributes when the @a2@ attribute  of
one of the objects has value @0@. The transitions have the 
 effect of altering
the attribute @a1@ of the  object @O@ and the attribute @a2@ of the
object @O'@.  ``Irrelevant'' attributes (such as @a3@ and 
@a2@ of @O@, and the \emph{right-hand side} occurrence of @a1@ of @O'@)
 need not be mentioned in a rule.

A \emph{subclass} inherits all the attributes and rules of its 
superclasses.

A Real-Time Maude specification is \emph{executable} under reasonable
conditions, 
and the tool offers a variety of formal analysis 
methods. The \emph{rewrite} command simulates \emph{one} fair 
behavior of the system \emph{up to a certain duration}. It is
written with syntax @(trew @$t$@ in time <= @$\tau$@ .)@, 
where $t$ is the initial state and
$\tau$ is a term of sort @Time@. 
The \emph{search} command uses a breadth-first strategy to analyze all
possible behaviors of the system, by checking whether a state matching
a \emph{pattern} can be reached from the initial state such that a
given \emph{condition} is satisfied.  The \emph{timed search} command, having the syntax

\myttsize
\begin{alltt}
  (tsearch [1] \(t\) =>* \(pattern\) such that \(cond\) in time <= \(\tau\) .)
\end{alltt}
\normalsize

\noindent works similarly, but  restricts the search to states reachable from the
initial state within time $\tau$.

Real-Time Maude  also extends Maude's \emph{linear temporal logic model
  checker} 
  to check whether
each behavior, possibly  up to a certain time bound,
  satisfies a linear temporal logic 
  formula.
 \emph{State propositions} are terms of sort @Prop@, and their 
semantics should be 
given by (possibly conditional) equations
 of the form @{@$\mathit{statePattern}$@}@$\,$@|=@$\;\,\mathit{prop}\;\,$@=@$\,\;b$,  
for $b$ a term of sort @Bool@, which defines the state proposition 
$prop$ to hold in all
states $@{@t@}@$ where $@{@t@}@$ \verb+|=+ $prop$ evaluates to @true@.
A temporal logic \emph{formula} is constructed by state
propositions and
temporal logic operators such as @True@, @False@, @~@ (negation),
@/\@, @\/@, @->@ (implication), @[]@ (``always''), @<>@
(``eventually''), and @U@ (``until'').
The time-bounded model checking command has syntax 
\begin{alltt}
  (mc \(t\) |=t \(\mathit{formula}\) in time <= \(\tau\) .)
\end{alltt}
for initial state $t$ and temporal logic formula $\mathit{formula}$ .

 Finally, the
\verb@find earliest@ command determines the shortest time needed to
reach a desired state.

For time-nondeterministic tick rules (i.e., tick rules in which the matching substitution 
does not uniquely determine the duration of the tick), the above model checking commands are applied
according to the chosen time sampling strategy, so that only a subset of all possible
behaviors is analyzed.

\section{Thermal Systems}\label{sec:thermalsys}

We define a \emph{thermal system} as a system whose
components behave according to the laws of physics and that are
related to heat transfer and temperature change.  \emph{Heat} is the
form of energy that can be transferred from one system to another as a
result of temperature difference.  The transfer of energy as heat is
always from the higher-temperature medium to the lower-temperature
one. The flow of heat  to a component  effects the component by  either 
  changing its  temperature or by
changing  the component's phase (such as ice melting to water).


\paragraph{Temperature change.} The
equation representing the heat gained or lost by an object as it
undergoes a temperature change $\Delta T$ is given by $\Delta Q = m
\cdot c \cdot \Delta T$, where $m$ is the mass of the object and $c$
is the specific heat of the object at constant volume.  The amount of
heat transferred per time unit, the \emph{heat transfer rate}, can be
described as $\dot{Q} = m \cdot c \cdot \dot{T}$, where $\dot{T}$
represents the change of temperature per time unit.

\par
Heat can be transferred by three different mechanisms:
\emph{conduction}, \textit{convection}, and \textit{radiation}.
\begin{itemize}
\item \textit{Conduction}
occurs when heat flows through stationary materials. The rate of heat
conduction through a medium depends on the geometry of the medium, the
thickness of the medium, the material of the medium, and temperature
difference across the medium.  The \textit{conduction rate} is given
by $\dot{Q} = \frac{k \cdot A}{L} \cdot (T_{1} - T_{2})$, where $k$ is
the thermal conductivity of the material, $A$ is the area of the
conduction, $L$ is the thickness of the material through which the
conduction occurs, and $T_{1}$, $T_{2}$ are the temperatures of the two
different media.  
\item \emph{Convection}
occurs when a moving fluid transports heat from a hotter object to a
colder object.  The rate of heat convection is proportional to the
temperature difference, and is expressed by Newton's law of cooling as
$\dot{Q} = h \cdot A \cdot (T_{1} 
- T_{2})$, where $h$ is the convection heat transfer coefficient, 
and $A$ is the surface area through which convection heat transfer
takes place. 
\item \emph{Radiation}
  occurs directly through the components' surface, without any
  transfer medium.  The rate of heat radiation is given by $\dot{Q} =
  \varepsilon \cdot \sigma \cdot A \cdot (T_{1}^{4} - T_{2}^{4})$,
  where $\varepsilon$ is the emissivity of the surface, $\sigma$ is
  the Stefan-Boltzmann constant, and $A$ is the surface area through
  which radiation heat transfer takes place.
\end{itemize}

For a running example, consider a cup of hot coffee in a room.  If the
temperature of the coffee is higher than that of the room, heat
will  flow  from the coffee to the room, until both of them reach the
same temperature.  In particular, the heat flows
 through the cup wall by conduction and radiation, and from
the surface of the coffee (assuming there is no lid on the cup) to the
room by convection. 

\paragraph{Phase transition.} 
Another important aspect that must be taken into account when modeling
thermal systems is the different \emph{phases} of a substance.  For
example, water has three distinct phases: \emph{solid} (ice),
\emph{liquid}, and \emph{gas} (vapor).  The \emph{latent heat} is the
amount of energy released or absorbed by a chemical substance during a
phase transition. Note that during a \emph{phase transition} the
temperature does not change; the flow of energy goes in as the latent
heat of the transition process.  For example, to change its phase from
solid to liquid, water goes through the melting process which consumes
some energy.

\section{A Framework for Modeling Physical Systems}\label{sec:modeling_framework}

This section presents a high-level overview of the general method we
use to model physical systems and to approximate their behavior. This
method adapts the effort-and-flow-variable approach common in the
field of physical systems modeling (see, e.g.,~\cite{wellstead}). In
particular, Section~\ref{sec:ps-comp} briefly describes the modeling
of physical systems in this framework,
and
Section~\ref{sec:model-thermal} shows how our method can be
instantiated to model thermal systems. 

\subsection{Modeling Physical Systems}\label{sec:ps-comp}

A well known modeling approach for physical systems is based on two
kinds of variables in the model specification: \emph{effort} 
and \emph{flow} 
variables~\cite{wellstead}.

The components of a physical system can be thought of as energy
manipulators which process the energy injected into the system.  These
energy manipulators interact through energy ports.  For this purpose,
effort and flow variables are used to represent the power being
transmitted through energy ports.  The flow variable is associated
with the act of delivering energy, whereas the effort variable is
associated with the act of measuring the flow of energy. For example,
our coffee system could have two effort variables, one denoting the
temperature of the coffee and one denoting the temperature of the
room, and one flow variable, denoting the flow of heat from the coffee
to the room per time unit.

The approach using effort/flow  variables is applicable to
different areas of physical systems.  In mechanical translation
systems, the pair of effort and flow variables are force and velocity;
in mechanical rotation systems, torque and angular velocity; in
electrical systems, voltage and current; in fluidic systems, pressure
and volume flow rate; in thermal systems, temperature and heat flow
rate~\cite{bentley}. 

\begin{figure}[t]
\centering
\includegraphics[width=0.7\textwidth]{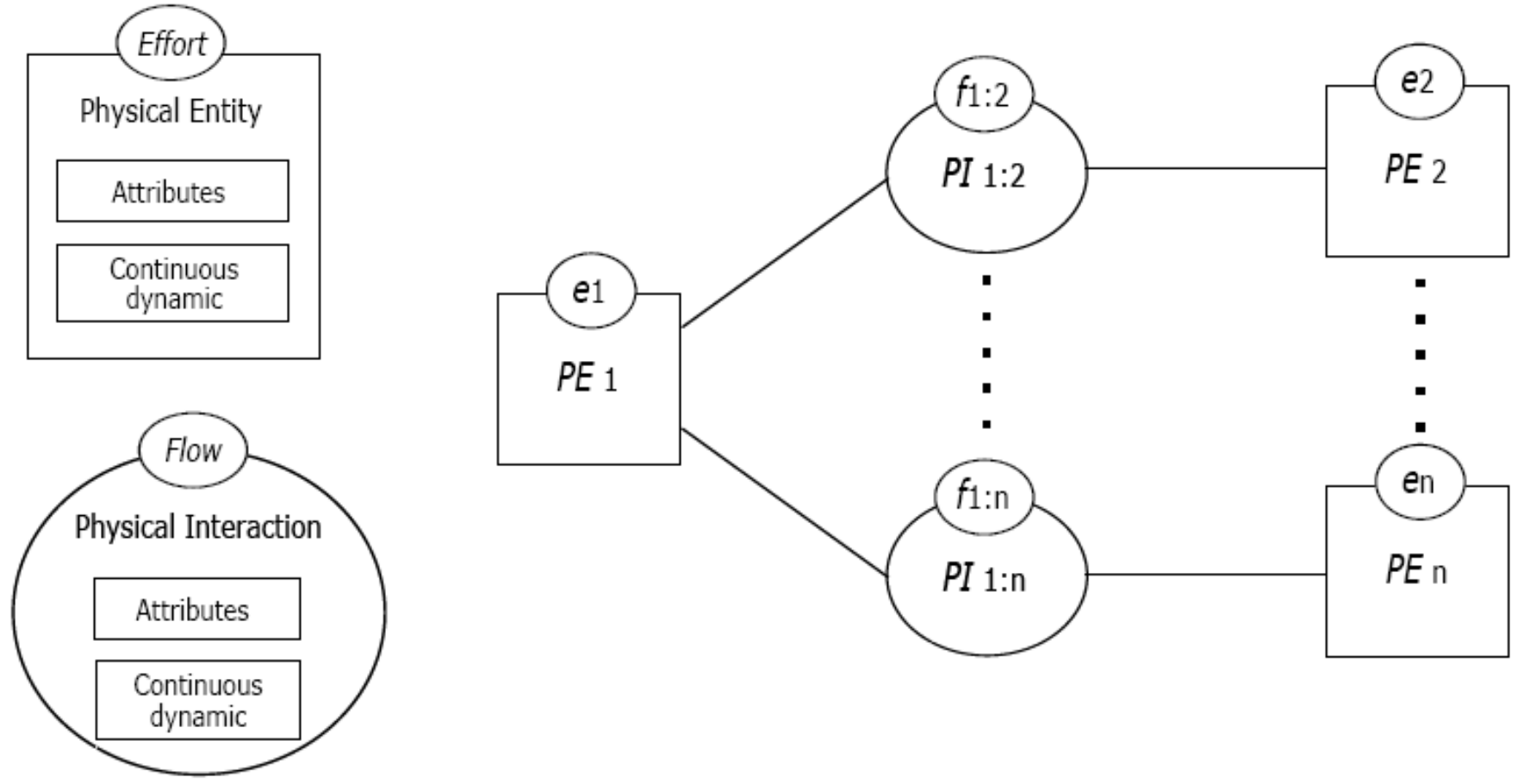}
\caption{Physical system components.}
\label{fig:ps_components}
\end{figure}

Our idea is to think of a physical system as the combination of two
kinds of components: \textit{physical entities} and \textit{physical
  interactions}, as shown in Fig.~\ref{fig:ps_components}.

A \emph{physical entity} (such as a cup of coffee) consists of a set
of \emph{attributes} and a \emph{continuous dynamics}.  We consider
three kinds of attributes: \emph{continuous} variables (denoting
physical quantities, such as the temperature, that change with time),
\emph{discrete} variables, and \emph{constants}.  A physical entity
has one \emph{effort variable}, which is a continuous variable and the
``main'' attribute of the physical entity.  The \emph{continuous
  dynamics} of a physical entity is given by a differential equation
which is used to compute the value of the effort variable. A physical
entity can have one or more physical interactions with one or more
physical entities.  The values of the flow variables of these
interactions are used in the computation of the continuous dynamics of
the physical entity.

A \emph{physical interaction} (such as the heat flow from the coffee
to the room) represents an interaction between two physical entities.
It consists of one \emph{flow variable}, a set of \emph{attributes},
and a \emph{continuous dynamics}.  The flow variable is a continuous
variable.  The continuous dynamics is described by a differential
equation.  The values of the effort variables from the two physical
entities are used in the computation of the continuous dynamics of the
physical interaction.

\subsubsection{Physical System Behavior}\label{sub2sec:ps_beh}

\begin{figure}[t]
\centering
\includegraphics[width=0.8\textwidth]{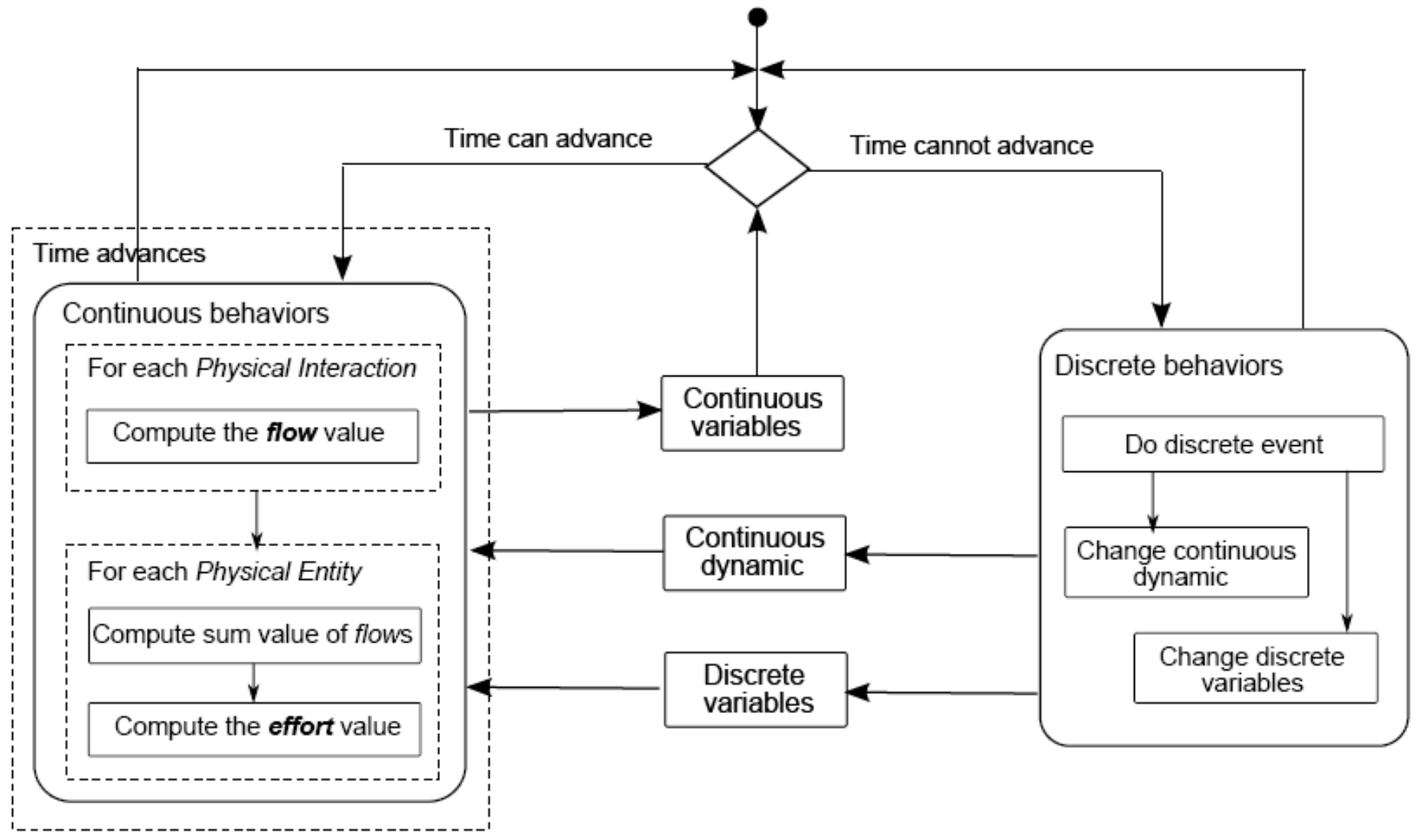}
\caption{Physical system behaviors.}
\label{fig:ps_behaviors}
\end{figure}

The basic behavior of physical system components is their continuous
behavior.  However, there are natural phenomena that exhibit
 a combination of continuous and discrete behaviors.  
In our method, the continuous behavior is performed as
long as  time advances.  When  a discrete event must happen,
  time cannot advance until the event has been performed.
  Figure~\ref{fig:ps_behaviors}
shows the execution of physical system behaviors. 

\paragraph{Approximating continuous behaviors.}
We use a numerical approach to approximate the continuous behaviors of the
physical system components by advancing time in discrete time steps,
and computing the values of the continuous variables at each 
``visited'' point in time. 
There are two main steps in the computation of the continuous behavior
which is performed in each time step.  The first step is to compute
the new value of the flow 
variable of each physical interaction in the system.  This computation
is based on the continuous dynamics of the physical interaction, the
values of its attributes,  and the 
 ``old'' values (i.e., the values computed at the previous visited point in
 time)  of the effort variables of the two
physical entities connected to the interaction.   In the configuration of physical
system components in Figure~\ref{fig:ps_components}, the new values of
the flow variables $f_{1:2}$ to $f_{1:n}$ of the physical interactions
$PI_{1:2}$ to $PI_{1:n}$, respectively, are computed. The computation
of the new value of  $f_{1:2}$ of $PI_{1:2}$, for example,  is not only
based on the attributes of the physical interaction, but also on the
``old''  values of
$e_{1}$ of $PE_{1}$ and $e_{2}$ of $PE_{2}$. 

The second step is to compute the ``new'' value of the effort variable of each
physical entity in the system.  Beside the current value of the effort
variable, the continuous dynamic of the physical entity, and the
values of its attributes, the computation also needs the sum  of the 
(new) values of the 
flow variables of all the physical interactions connected to it. 
For example, in the system in 
Fig.~\ref{fig:ps_components}, the new values of the effort variables $e_{1}$
to $e_{n}$ of the physical entities $PE_{1}$ to $PE_{n}$, respectively,
are computed.  The computation of the new value of $e_{1}$ of $PE_{1}$,
for example,  also depends on the sum  of the new values of 
$f_{1:2}$ to $f_{1:n}$.  

The difference between computing  the new values of the  effort
variables and the
 flow variables is that for the former we need a differential
equation solution technique, whereas for the latter we do not need
such a technique since a flow variable represents a rate. 

\paragraph{Discrete behaviors.}
The conditions for triggering discrete events depend on the values
of the continuous variables.  We consider
two kinds of discrete events:  one that changes the continuous dynamics
of a physical entity or interaction (e.g., when the coffee starts boiling, 
its continuous dynamics is changed), and 
one that changes the value of a discrete variable.  The change caused
by a discrete event will effect the computation of the continuous 
behavior in the next time step.

\subsection{Applying the Framework for  Modeling Thermal Systems}
\label{sec:model-thermal} 

\begin{figure}[htbp]
\centering
\includegraphics[width=0.7\textwidth]{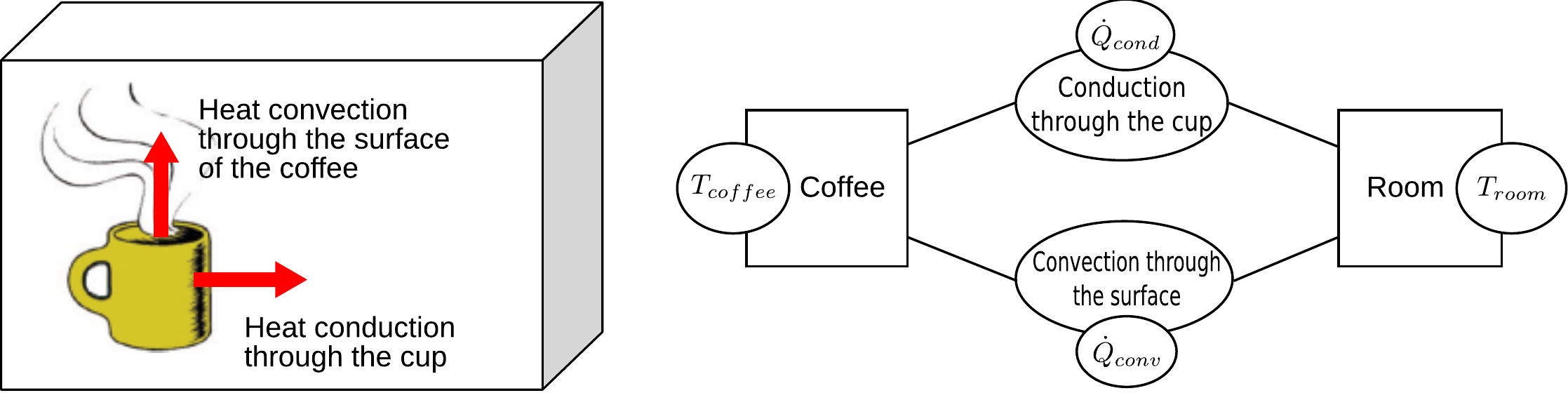}
\caption{A cup of coffee in a room with the convection and conduction heat transfers between them.}
\label{fig:example_coffee_room_conv_cond}
\end{figure}

We can instantiate the above framework to thermal systems as follows:   
a  physical entity is a  \emph{thermal entity}, whose effort
variable ($T$) 
denotes the temperature of the entity and whose continuous dynamics is
a differential  equation  defining the heat gained or lost by the entity as its temperature changes.
Likewise, a physical interaction is a \emph{thermal interaction} whose
flow variable ($\dot{Q}$) denotes the heat flow rate. We may have different kinds
of thermal interactions: conduction, convection, and radiation; their
continuous dynamics are given by differential equations used for the
heat transfer rates.    Table~\ref{tab:thermalsys_components} shows the thermal system
 components, and Fig.~\ref{fig:example_coffee_room_conv_cond} shows a model of our
example
with a hot cup of coffee in a room.

\begin{table}[ht]
\centering \ra{1.15} \caption{Thermal system components.}
    \begin{minipage}{\textwidth}
    \centering
    \resizebox{0.8\textwidth}{!}{
    \begin{tabular}{l c l c l c l}
    \toprule
    \textbf{Components} & \phantom{m} & \textbf{Effort/Flow} & \phantom{m} & \textbf{Attributes}  & \phantom{m} & \textbf{Continuous dynamic} \\
    \midrule
    Thermal entity && Effort: $T$ && $m$, $c$ && $\dot{T} = \frac{\Sigma{\dot{Q}}}{m \cdot c}$ \\
    Thermal interaction: conduction && Flow: $\dot{Q}$ && $A$, $k$, $L$ && $\dot{Q} = \frac{k \cdot A}{L} \cdot (T_{1} - T_{2})$ \\
    Thermal interaction: convection && Flow: $\dot{Q}$ && $A$, $h$ && $\dot{Q} = h \cdot A \cdot (T_{1} - T_{2})$ \\
    Thermal interaction: radiation && Flow: $\dot{Q}$ && $A$, $\varepsilon$, $\sigma$  && $\dot{Q} = \varepsilon \cdot \sigma \cdot A \cdot (T_{1}^{4} - T_{2}^{4})$ \\
     \bottomrule
    \end{tabular}
    }
    \end{minipage}
    \label{tab:thermalsys_components}
\end{table}

\section{Modeling Thermal Systems in Real-Time Maude}\label{sec:modeling_thermalsys}

This section explains how we can apply the general methods for
modeling physical systems given in
Section~\ref{sec:modeling_framework} to model and execute thermal
systems in Real-Time Maude. In particular,
Section~\ref{sec:thermal-classes} defines classes for thermal entities
and interactions, Section~\ref{sec:thermal-euler} explains how the
Euler method (see, e.g.,~\cite{hoffman}) can be used to define in
Real-Time Maude an approximation of the continuous dynamics of a
thermal system, Section~\ref{sec:thermal-hybrid} describes how hybrid
behaviors of thermal systems can be modeled, and
Section~\ref{sec:heat-gen} explains how external heat sources can be
easily added to our models.


\subsection{Defining Thermal System  Components}\label{sec:thermal-classes}

In our method, thermal entities (physical entities in thermal systems)
can be defined in Real-Time Maude as object instances of the following
class @ThermalEntity@:

\myttsize
\begin{alltt}
class ThermalEntity | heatCap : PosRat,   mass : PosRat, 
                      temp : Rat,         tempDisplay : String, 
                      heatTrans : Rat,    mode : CompMode .

sort CompMode .     op default : -> CompMode [ctor] .
\end{alltt}

\normalsize \noindent The attribute \texttt{heatCap} denotes the heat
capacity and \texttt{mass} the mass of the entity; \texttt{temp}
denotes the temperature, which corresponds to the effort variable in
thermal systems; \texttt{tempDisplay} is used to display the
temperature in a more readable format than as a rational number;
finally, \texttt{heatTrans} and \texttt{mode} 
are needed, as we will see later, to implement phase transitions,
i.e., to encode the computation mode of a thermal entity, which is
related to its discrete behavior.

We define a class for general \textit{thermal interactions}
(physical interactions in thermal systems), and three subclasses for
the three different heat transfer mechanisms:

\myttsize
\begin{alltt}
class ThermalInteraction | entity1 : Oid,   entity2 : Oid,   area : PosRat, 
                           Qdot : Rat,      QdotDisplay : String .

class Conduction | thermCond : PosRat, thickness : PosRat .
class Convection | convCoeff : PosRat .
class Radiation | emmissiv : PosRat .

subclass Conduction Convection Radiation < ThermalInteraction .
\end{alltt}

\normalsize \noindent In the \texttt{ThermalInteraction} class,
\texttt{Qdot} is the attribute corresponding to the heat flow rate
$\dot{Q}$ of the thermal interaction.  \texttt{QdotDisplay} is used to
display the heat flow rate in a readable format.  \texttt{entity1} and
\texttt{entity2} are object identifiers of the two thermal entities
connected by an object of this class, and \texttt{area} is the area of
the interaction common for all three interaction types. The remaining
interaction-specific attributes, e.g., for conductivity the thermal
conductivity \texttt{thermCond} of the material and the thickness
\texttt{thickness} of the material through which the conduction
occurs, are specified in the corresponding subclasses.


\subsection{Approximating the Continuous Behaviors using the Euler
  Method}\label{sec:basic-cont}\label{sec:thermal-euler}

We use ordinary differential equations for specifying the continuous
dynamics of physical entities and interactions.  For executing the
continuous behavior of a physical system by using a discrete-time
sampling strategy, we need an approximation to compute the values of
the effort and flow variables.  In this paper we adapt the Euler
method \cite{hoffman}, a numerical approach to solve differential
equations.  Assume that the value of the continuous variable $y$ at
time $t$ is defined by the differential equation $\dot{y}=f(y,t)$.
Assume furthermore that the value of $y$ at time  $t_n$ is $y_n$.
When time advances, the Euler method uses the function $y_{n + 1} =
y_{n} + h \cdot f(y_{n}, t_n)$ to compute the value of $y$ at time
point $t_{n+1}=t_n+h$, where $h$ is the time step.

To implement the Euler method, we define the
following functions :
\begin{itemize}
\item \texttt{computeQdot} that computes the  heat flow rate
$\dot{Q}$ of each thermal interaction,
\item \texttt{computeTemp} that computes the  temperature $T$ of
each thermal entity in the system,
\item \texttt{sumQdots} that computes the sum of the $\dot{Q}$s of the
  thermal interactions of each entity, and
\item \texttt{ODESol} that computes the next value $y_{n+1}=
  y_n+h*f(y_n,t_n)$ of a continuous variable $y$ using the Euler
  method.
\end{itemize}
We start with the declaration of the following variables:
\myttsize
\begin{alltt}
vars T T1 T2 QDOT YN FYN QT NEW-Qdot CURR-TEMP NEW-HEAT-TRANS : Rat .
vars k A L EPSILON c m h : PosRat .    var S : OnOff .
vars E E1 E2 TI HG : Oid .             vars CONFIG REST : Configuration .
\end{alltt}
\normalsize

The function \texttt{computeQdot} computes the  heat flow rate
$\dot{Q}$ of each thermal interaction in the system according to the
laws given in Section~\ref{sec:thermalsys} and
Table~\ref{tab:thermalsys_components}. The following equation computes
the @Qdot@ (given by $\dot{Q}=\frac{k\cdot A}{L} \cdot (T_1 - T_2)$ as
explained in Section~\ref{sec:thermalsys}) of a thermal
\emph{conduction}
 @TI@ between two thermal entities
@E1@ and @E2@, and then recursively applies the @computeQdot@ function on
the remaining configuration: 

\myttsize
\begin{alltt}
op computeQdot : Configuration ~> Configuration .
ceq computeQdot(
   < E1\! :\! ThermalEntity | temp : T1 >   
   < E2\! :\! ThermalEntity | temp : T2 >
   < TI\! :\! Conduction\! |\! entity1\!\! :\! E1, entity2\!\! :\! E2, thermCond\!\! :\! k, area\!\! :\! A, thickness\!\! :\! L\! > 
   REST)
  =
   < TI\! :\! Conduction | Qdot\! : \emph{NEW-Qdot}, QdotDisplay\! : display(\emph{NEW-Qdot}, precision) >
   computeQdot(<\! E1\! :\! ThermalEntity\! |\! >  <\! E2\! :\! ThermalEntity\! |\! >  REST)
if \emph{NEW-Qdot := (k * A * (T1 - T2)) / L} .
\end{alltt}
\normalsize

 Likewise, the following two equations compute the new value of @Qdot@
 of, respectively, thermal \emph{convections} and \emph{radiations},
 and the last equation removes the  @computeQdot@ operator when there
 are no more  thermal interactions in the remaining configuration:

\myttsize
\begin{alltt}
ceq computeQdot(
   < E1\! :\! ThermalEntity | temp\! : T1 >
   < E2\! :\! ThermalEntity | temp\! : T2 >
   < TI\! :\! Convection | entity1\! : E1, entity2\! : E2, convCoeff\! : h, area\! : A > REST) 
  =
   < TI\! :\! Convection | Qdot\! :\! \emph{NEW-Qdot}, QdotDisplay\! :\! display(\emph{NEW-Qdot}, precision) >
   computeQdot(< E1\! :\! ThermalEntity\! |\! >  <\! E2\! :\! ThermalEntity\! |\! >  REST)
if \emph{NEW-Qdot := h * A * (T1 - T2)} .

ceq computeQdot(
   < E1\! :\! ThermalEntity | temp\! :\! T1 >
   < E2\! :\! ThermalEntity | temp\! :\! T2 >
   < TI\! :\! Radiation | entity1\! :\! E1, entity2\! :\! E2, emmissiv\! :\! EPSILON, area\! :\! A > REST) 
  =
   < TI\! :\! Radiation | Qdot\! :\! \emph{NEW-Qdot}, QdotDisplay\! :\! display(\emph{NEW-Qdot}, precision) >
   computeQdot(< E1\! :\! ThermalEntity | >  < E2\! :\! ThermalEntity\! |\! >  REST)
if \emph{NEW-Qdot := EPSILON * stefBolzConst * A * ((T1\! ^\! 4) - (T2\! ^\! 4))} .

eq computeQdot(CONFIG) = CONFIG [owise] .
\end{alltt}
\normalsize

The function \texttt{computeTemp} computes the  temperature $T$ of
each thermal entity in the system, with the dynamics  given by
$\dot{T} = \frac{\sum \dot{Q}}{m\cdot c}$, where $\sum \dot{Q}$ is the
sum of the $\dot{Q}$s of the thermal interactions of the entity which 
 is computed by the function @sumQdots@ below. The function 
uses a numerical
method (\texttt{ODESol})  which  needs the value of the
current temperature (\texttt{T}), the time step used, and
the continuous dynamics: 

\myttsize
\begin{alltt}
op computeTemp\! :\! Configuration ~> Configuration .
ceq computeTemp(<\! E\! :\! ThermalEntity\! |\! heatCap\!\!\! :\!\! C, mass\!\! :\! M, temp\!\! :\! T, mode\!\! :\! default\! >\, REST) 
    =
    <\! E\! :\! ThermalEntity\! |\! temp\!\! :\! \emph{CURR-TEMP}, tempDisplay\!\! :\! display(\emph{CURR-TEMP}\!,\! precision)\! > 
    computeTemp(REST)
   if \emph{CURR-TEMP := ODESol(T, timeStep, sumQdots(REST,\! E)\! /\! (M\! *\! C)))} .

eq computeTemp(CONFIG) = CONFIG [owise] .
\end{alltt}
\normalsize

The function \texttt{sumQdots} computes the sum values of the heat flow rate of all thermal interactions
connected to a thermal entity. The direction of the heat flow determines whether a thermal
entity gains or loses the heat.  This is represented by multiplying
the heat flow rate value by minus one for one part of the
interaction:

\myttsize
\begin{alltt}
op sumQdots\! :\! Configuration Oid  ~> Rat .
eq sumQdots(<\! TI\! :\! ThermalInteraction\! |\! entity1\!\! :\! E1, entity2\!\! :\! E2, Qdot\!\! :\! QDOT\! >\!  REST, E) 
   = 
   if (E == E1 or E == E2) then
      (if E == E1 then -1 * QDOT + sumQdots(REST, E) else QDOT + sumQdots(REST, E) fi)
   else sumQdots(REST, E) fi .

eq sumQdots(CONFIG, E) = 0 [owise] .
\end{alltt}
\normalsize

The function \texttt{ODESol} uses the Euler method to compute
the next value $y_{n+1}= y_n+h*f(y_n,t_n)$ of a continuous variable $y$,  given the
current value $y_n$ of
the variable, the time step $h$, and the value $f(y_n, t_n)$ computed from the
continuous dynamics of the variable:

\myttsize
\begin{alltt}
op ODESol : Rat PosRat Rat -> Rat .
eq ODESol(YN, H, FYN) = YN + H  * FYN .
\end{alltt}
\normalsize

As explained in Section~\ref{sec:rtm}, a \textit{tick rule} models the
advance of time in a system. We define the following tick rule that
advances time by one time step and computes the new temperatures for
all entities:

\myttsize
\begin{alltt}
rl [tick]\! :\! \texttt{\char123}CONFIG\texttt{\char125} => \texttt{\char123}computeTemp(computeQdot(CONFIG),timeStep))\texttt{\char125} in time timeStep .
\end{alltt}
\normalsize


\subsection{Defining Thermal System with Hybrid Behaviors}\label{sec:thermal-hybrid}

As mentioned in Section~\ref{sub2sec:ps_beh}, a physical system may
also exhibit (at least) two kinds of discrete behaviors. In this paper
we focus on discrete events that change the continuous dynamics of an
entity or interaction. In thermal systems, we have \emph{phase
  transitions} between different phases of the entities. For the
example of water, a discrete event changes the phase from ``solid'' to
``melting,'' thereby also changing the continuous dynamics of the
water: the received thermal energy is used for the melting
process and the temperature of the water does not change during
melting. The following changes need to be made to our model to
accommodate such hybrid behaviors in physical systems:
\begin{enumerate}
\item The basic @ThermalEntity@ class must be extended to keep track
  of the phases of the entity.
\item Instantaneous rewrite rules modeling the discrete change from
  one phase to another must be added.
\item It must be ensured that the above rules are applied at the right
  time. 
\item The functions computing the values of the continuous variables
  must take into account the different continuous dynamics in
  different phases.
\end{enumerate}

\begin{figure}[htbp]
\centering
\includegraphics[width=0.8\textwidth]{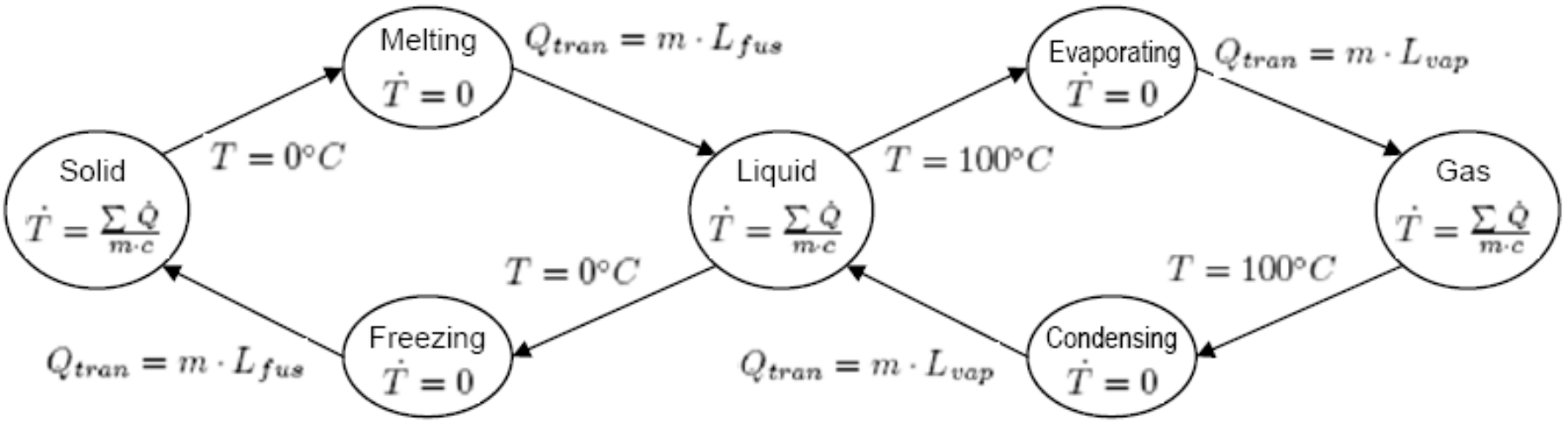}
\caption{The phases  of the water entity.}
\label{fig:phasetrans_water}
\end{figure}

\paragraph{Extending the basic class.}
  Let's say we want to define a thermal
entity representing a water substance with three phases: solid,
liquid, and gas.  Between these main phases, the water is
either melting, freezing, evaporating, or condensing. Figure~\ref{fig:phasetrans_water} 
shows the various phases (and their corresponding continuous dynamics)
and the transitions between them.  

The class @WaterEntity@ extending the base class @ThermalEntity@ can
be defined as follows:

\myttsize
\begin{alltt}
class WaterEntity | phase\! :\! MatterState,  heatTrans\! :\! Rat,  heatTransDisplay\! :\! String .
subclass WaterEntity < ThermalEntity .

sort MatterState .
ops liquid solid gas melting evaporating condensing freezing\! :\! -> MatterState .
op phaseChange\! :\! -> CompMode .
\end{alltt}
\normalsize 

\noindent The new attribute \texttt{phase} denotes the current  phase
 of the water substance, and \texttt{heatTrans} denotes 
the  latent heat of the water in the transitions between the three
main phases.  Notice that the continuous dynamics of the temperature is the same in all the
three main phases of water, whereas the temperature does not change
in-between these phases. In addition to the @default@ computation mode
defined above, we also add a new computation mode @phaseChange@ to
denote these intermediate phases of water, so that the @computeTemp@
function is defined correctly.

\paragraph{Rewrite rules defining discrete change.}  
A phase change is modeled by an instantaneous rewrite rule. We
show two of the eight such rules (corresponding to the
arrows in Fig.~\ref{fig:phasetrans_water}) for water:

\myttsize
\begin{alltt}
crl [solid-to-melting]\! :\! 
    < E\! :\! WaterEntity | temp\! :\! T, phase\! :\! \emph{solid} > 
   => 
    < E\! :\! WaterEntity | phase\! :\! \emph{melting}, mode\! :\! phaseChange, heatTrans\! :\! 0 > 
   if T >= 0 .

crl [melting-to-liquid]\! :\! 
    < E\! :\! WaterEntity | phase\! :\! \emph{melting}, heatTrans\! :\! QT, mass\! :\!  m > 
   => 
    < E\! :\! WaterEntity | phase\! :\! \emph{liquid}, mode\! :\! default > 
   if QT / m >= latentHeatFus .
\end{alltt}

\normalsize \noindent The change from a ``main'' phase  to a
``transitional'' phase
  happens when  the temperature reaches a given  value, whereas a
  transition the
  other way happens when  the value of the heat accumulated during the
  transitional phase
 divided by the mass of the  entity reaches some value
(such as the latent heat fusion).  

We ensure that the above rules are applied when they should by
disabling the tick rule when one of the above rules is enabled. This
is done by defining a function @timeCanAdvance@ on the state, so that 
@timeCanAdvance@ holds only if no such instantaneous rule needs to be
applied in the current state. We then modify the tick rule so that it
can only be applied to states where @timeCanAdvance@ holds. Again, we
only show two of the equations defining @timeCanAdvance@ and the
modified tick rule:

\myttsize
\begin{alltt}
eq timeCanAdvance(< E\! :\! WaterEntity | phase\! :\! solid, temp\! :\! T >) =  T < 0 .
eq timeCanAdvance(< E\! :\! WaterEntity | phase\! :\! melting, heatTrans\! :\! QT, mass\! :\!  m >) =
      QT / m < latentHeatFus .

crl [tick]\! :\! \char123CONFIG\char125 => \char123computeTemp(computeQdot(CONFIG), timeStep)\char125 in time timeStep 
            if \emph{timeCanAdvance(CONFIG)} .
\end{alltt}
\normalsize

The change of continuous dynamics caused by a  phase change  means
that there is more than one possible continuous dynamics for a
thermal system component. We need to modify the function used to
compute the corresponding continuous variable. For the water entity,
 we need to modify the function \texttt{computeTemp}, so that when the
 entity is in mode @phaseChange@ (as opposed to the @default@ mode,
 where the equation in Section~\ref{sec:basic-cont} applies), then the
 temperature does not change, but the accumulated heat is increased/decreased:

%

\myttsize
\begin{alltt}
ceq computeTemp(< E\! :\! WaterEntity | mode\! :\! \emph{phaseTrans}, heatTrans\! :\! QT > REST) =
       < E\! :\! WaterEntity | heatTrans\! :\! \emph{NEW-HEAT-TRANS}, 
                          heatTransDisplay\! :\! display(\emph{NEW-HEAT-TRANS}, precision) > 
       computeTemp(REST)
   if \emph{NEW-HEAT-TRANS := ODESol(QT, timeStep, sumQdots(REST, E))} .
\end{alltt}
\normalsize


\subsection{External Heat Sources}\label{sec:ts_rtm_heatgen}\label{sec:heat-gen}

It is sometimes convenient to be able to inject heat into a thermal entity. For example, we may want
to add a boiler to our coffee system that adds heat to the cup of coffee. 
 In our method, such heat generators can be defined in 
Real-Time Maude as object instances of the following class @HeatGenerator@:

\myttsize
\begin{alltt}
class HeatGenerator | Qdot : Rat, entity : Oid .
\end{alltt}
\normalsize

\noindent The attribute @Qdot@ denotes the heat flow rate provided by the heat generator; 
@entity@ is the object identifier of the thermal entity connected to the heat generator. 

Adding a heat generator to a thermal entity means that beside the heat flows 
from the interactions with other thermal entities, the heat flow 
from the heat generator influences the change of the temperature of the entity.  
We need to modify the function used to compute the sum of heat flow 
rate values of a thermal entity to include the heat flow rate from the 
connected heat generator:

\myttsize
\begin{alltt}
eq SumQdots((< HG : \emph{HeatGenerator} | Qdot : QDOT, entity : E1 > REST), E) =
      if (E == E1) then QDOT + SumQdots(REST, E) else SumQdots(REST, E) fi .
\end{alltt}
\normalsize

\begin{table}[t]
\centering
\ra{1.3}
\caption{Initial values for object constant parameters ($\pi = \frac{22}{7}$).}

\begin{minipage}{\textwidth}
\resizebox{\textwidth}{!}{
\begin{tabular}{l l r l c l l r l }
\toprule 

\multicolumn{4}{l}{\textbf{room}} & \phantom{\hspace{3em}} & \multicolumn{4}{l}{\textbf{cup of coffee}}  \\
\midrule
\texttt{airDensity} & air density   \footnote{\scriptsize{At sea level and $20 ^\circ C$, air has a density of approximately $1.2 kg$ (http://en.wikipedia.org/wiki/Density\_of\_air). }}   & $\frac{12}{10}$       & $kg/m^3$ && \texttt{cupRadius} & inner radius     & $\frac{4}{100}$       & $m$ \\
\texttt{roomVolume} & volume        & \scriptsize{$64$}                 & $m^3$ && \texttt{cupHeight} & height      & $\frac{9}{100}$       & $m$ \\
\texttt{roomMass} & mass        & $\frac{384}{5}$       & $m^3$ && \texttt{cupCircum} & circumference   & $\frac{44}{175}$      & $m$ \\
\texttt{roomHC} & heat capacity & $\frac{105}{10}$      & $kJ/kg$   && \texttt{cupBaseArea} & circle area   & $\frac{22}{4375}$         & $m^2$ \\
\texttt{h} & h\footnote{\scriptsize{Free convection of gases is between $2$-$25$ $\frac{W}{m^2 \cdot ^\circ C}$ .}} & $\frac{20}{1000}$         & $\frac{kW}{m^2 \cdot ^\circ C}$ && \texttt{cupSideArea} & side area       & $\frac{99}{4375}$         & $m^2$ \\
&&&&& \texttt{cupThickness} & cup thickness  & $\frac{1}{200}$       & $m$ \\
&&&&& \texttt{waterDensity} & water density & \scriptsize{$1000$    }           & $kg/m^3$\\
&&&&& \texttt{coffeeVolume} & volume        & $\frac{99}{218750}$   & $m^2$ \\
&&&&& \texttt{coffeeMass} & mass        & $\frac{396}{875}$         & $kg$ \\
&&&&& \texttt{coffeeHC} & heat capacity     & $\frac{42}{10}$       & $kJ/kg$ \\
&&&&& \texttt{k} & k\footnote{\scriptsize{Thermal conductivity of porcelain}}       & $\frac{15}{10000}$        & $\frac{kW}{m \cdot ^\circ C}$ \\
\bottomrule
\end{tabular}
}
\par
   \vspace{-0.75\skip\footins}
   \renewcommand{\footnoterule}{}
\end{minipage}
\label{tbl:initParameters}
\end{table}%

\section{Case Studies} \label{sec:case-studies} This section shows how
our method can be used to model and analyze some simple thermal
systems. In particular, in Section~\ref{sec:case1} we model and
analyze a cup of hot coffee in a room, with two kinds of interactions
(\textit{conduction} and \textit{convection}). In
Section~\ref{sec:case2} we add a heater that sends constant heat to
the cup of coffee. Finally, in Section~\ref{sec:case3} we add a
``smart'' heater that switches itself \textit{on} or \textit{off}
in order to keep the temperature of the coffee in a desired interval.
It is worth noticing that, using the definitions in the previous
section, the definition of the first two models reduces to defining
appropriate initial states.  Table~\ref{tbl:initParameters} shows the
constants used for modeling the room and the cup of coffee.  The
analyses are performed on an \textit{AMD Athlon(tm) 64 Processor
  3500+} with 2GB of RAM, using a \texttt{timeStep} of $1$, unless
something else is specified.

\subsection{Case Study 1: A Cup of Coffee in a Room}
\label{sec:case1}

In our first case study we model the simple coffee system in
Fig.~\ref{fig:example_coffee_room_conv_cond}  in
Section~\ref{sec:model-thermal}  by defining the corresponding initial
state as follows: 

\myttsize
\begin{alltt}
op cs1 : -> GlobalSystem .
eq cs1 =
  \texttt{\char123}< coffee\! :\! ThermalEntity | heatCap\! :\! coffeeHC, mass\! :\! coffeeMass, heatFlow\! :\! 0,
                            \!\!  temp\! :\! 70, tempDisplay\! :\! "", mode\! :\! default, heatTrans\! :\! 0\! >
   < room\! :\! ThermalEntity | heatCap\! :\! roomHC, mass\! :\! roomMass, temp\! :\! 20, mode\! :\! default,
                     \!\!       tempDisplay\! :\! "", heatFlow\! :\! 0, heatTrans\! :\! 0\! >
   < crConduct\! :\! Conduction\! |\! entity1\! :\! coffee, entity2\! :\! room, thermCond\! :\! k, Qdot\! :\! 0,
              \!\!         \!\!       area\! :\! condArea, thickness\! :\! cupThickness\!, QdotDisplay\!\! :\! ""\! >
   < crConvect\! :\! Convection | entity1\! :\! coffee, entity2\! :\! room, convCoeff\! :\! h, 
                        \!\!      area\! :\! convArea, Qdot\! :\! 0, QdotDisplay\! :\! "" >\texttt{\char125} .
\end{alltt}
\normalsize 

\noindent where \texttt{coffeeHC, coffeeMass, \ldots} are constants of sort \texttt{Rat} 
with values as shown in Table~\ref{tbl:initParameters}. The constant
\texttt{condArea} is the conduction area, computed as the sum of the 
coffee circle area on the base of the cup and the coffee cylindrical 
area on the lateral surface of the cup.
The constant \texttt{convArea} is the convection area, equal 
to the coffee circle area, representing the top surface of the coffee.

We can simulate the behavior from this initial state 
up to a certain time, using Real-Time Maude's 
timed rewrite command:

\myttsize
\begin{alltt}
Maude> \textit{(trew cs1 in time <= 1000 .)}

result ClockedSystem :
  \texttt{\char123}< crConduct\! :\! Conduction | QdotDisplay\! :\! "0.0044820616", ... >
   < crConvect\! :\! Convection | QdotDisplay\! :\! "0.0000543280", ... >
   < coffee\! :\! ThermalEntity | TempDisplay\! :\! "21.6767974687", ... >
   < room\! :\! ThermalEntity | TempDisplay\! :\! "21.1390469168", ... >\texttt{\char125} in time 1000
\end{alltt}
\normalsize 

\noindent We show only the \textit{display}
attributes of the objects.
The simulation takes about a minute and a half.
The result is as expected: in $1000$ seconds, the temperature of the coffee is sensibly decreased,
reaching almost that of the room, whose temperature has 
increased slightly due to the heat from the coffee. 

\begin{figure}[htb]%
  \centering
  \subfloat[Case study 1: simulation]{\includegraphics[width=0.5\textwidth]{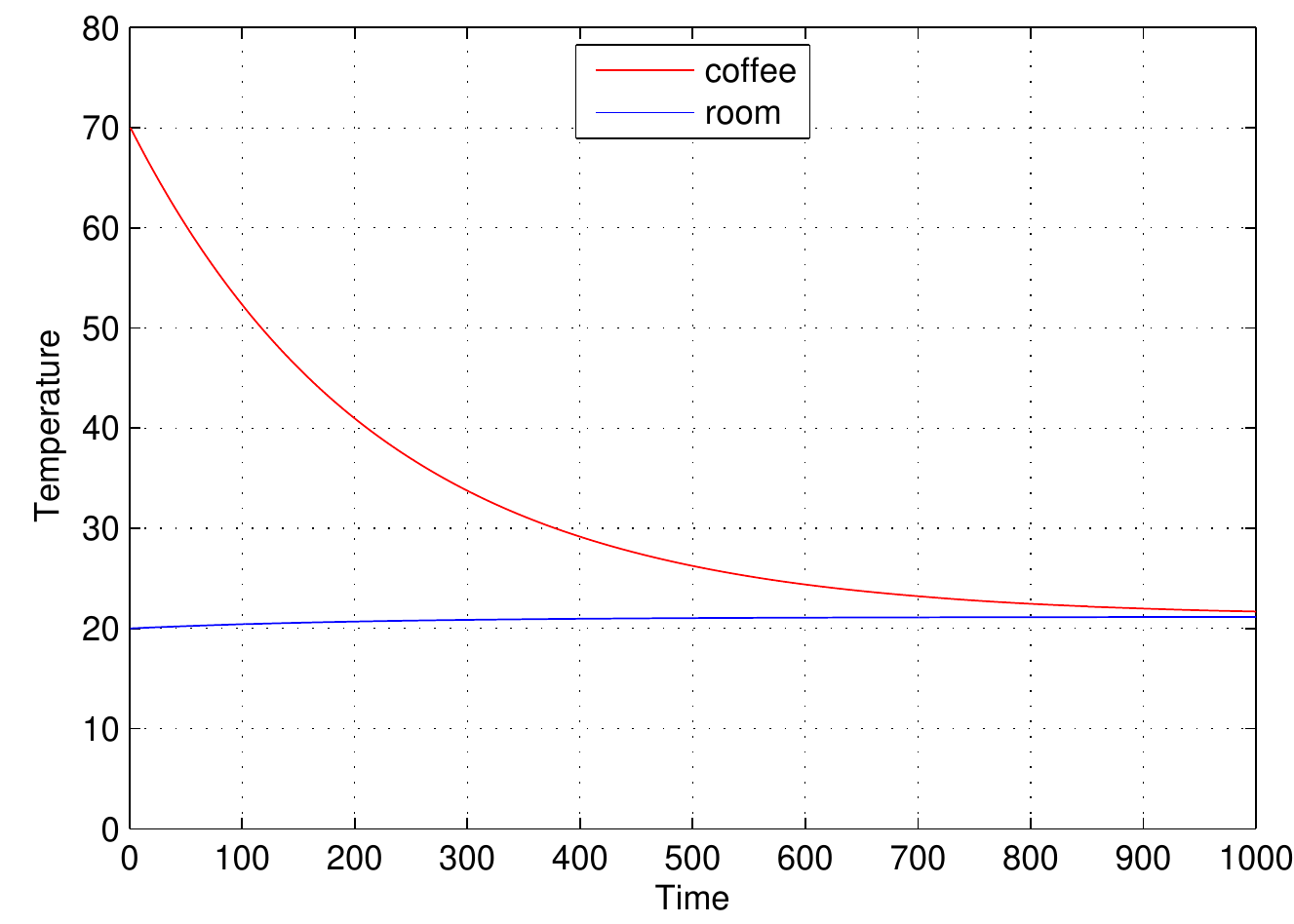}\label{fig:plotCS1}}%
  \subfloat[Case study 2: simulation]{\includegraphics[width=0.5\textwidth]{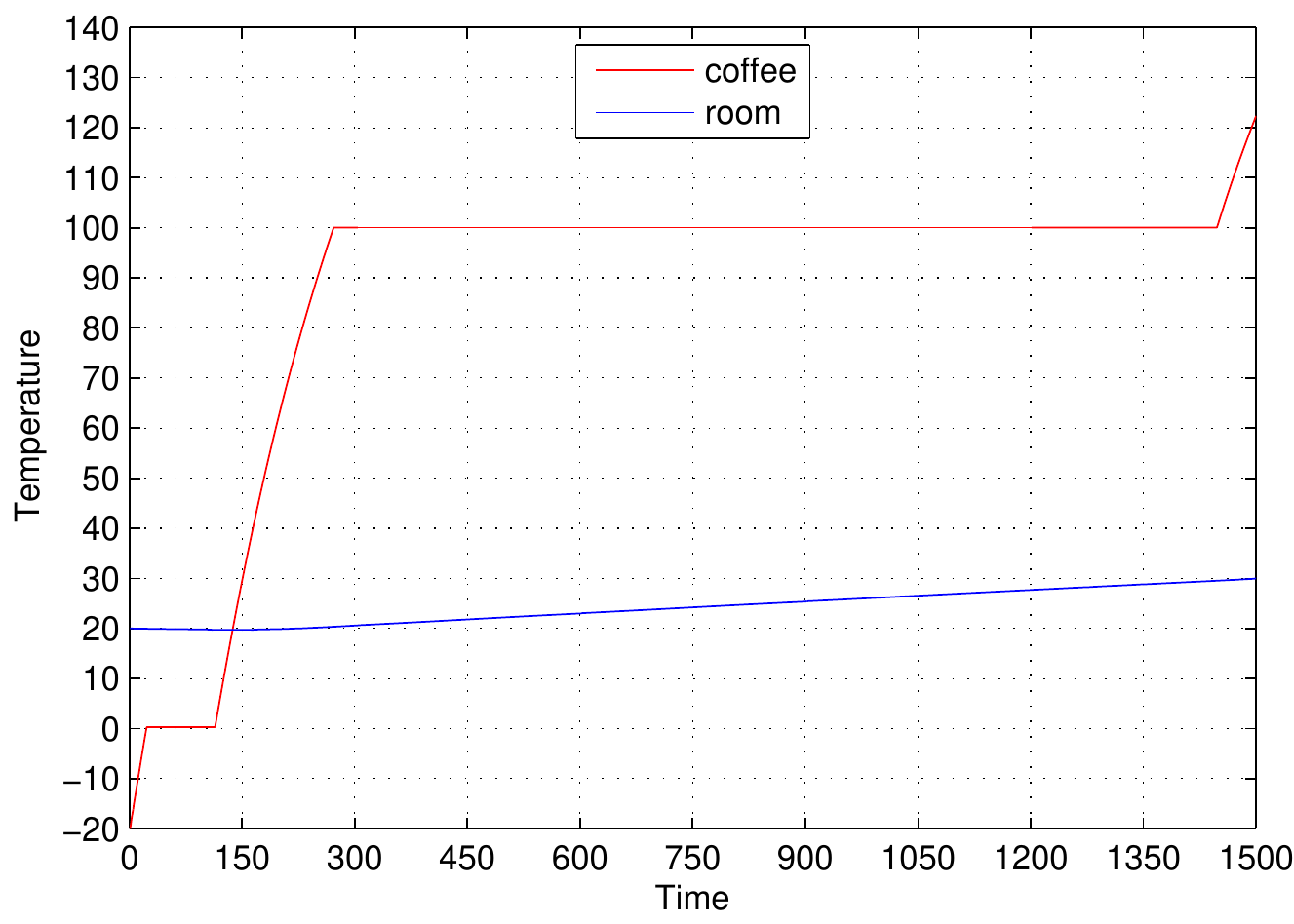}\label{fig:plotCS2}}
  \caption{Simulation plots for the case studies 1 and 2.}%
  \label{fig:plotCS1-CS2}%
\end{figure}

Moreover, we can collect the values for coffee and room temperatures
at each time step during the simulation.  This has been done by adding
a ``collector'' object to the initial configuration, which, at each
tick rule application, records the current time and temperature
values.  This allows us to plot the coffee and room temperatures as a
function over time, as shown in
Figure~\ref{fig:plotCS1-CS2}\subref{fig:plotCS1}.


Suppose we would like to check how long 
it takes for the coffee and room to reach about the same temperature, with a given tolerance.
This can be done by searching for one reachable state where the 
difference of the two temperatures is less than the given tolerance.
This kind of search can be performed in Real-Time Maude without any restriction on time,
since we can always reach such a desired state within a finite amount of time:

\myttsize
\begin{alltt}
Maude> \textit{(tsearch [1] cs1 =>* 
         \char123< coffee\! :\! ThermalEntity | temp\! :\! Tcoffee:Rat > 
          < room\! :\! ThermalEntity | temp\! :\! Troom:Rat >  C:Configuration\char125
             such that (abs(Tcoffee:Rat - Troom:Rat) <= 1/1000 ) with no time limit .)}
\end{alltt}
\normalsize 

\noindent It takes about $17$ minutes for Real-Time Maude to find a solution, which indicates that it takes 
around 2340 seconds for the room and coffee to reach the same temperature.


\subsection{Case Study 2: Adding a Boiler to the Cup of Coffee}
\label{sec:case2}


 \begin{figure}[t]
 \centering
 \includegraphics[width=0.5\textwidth]{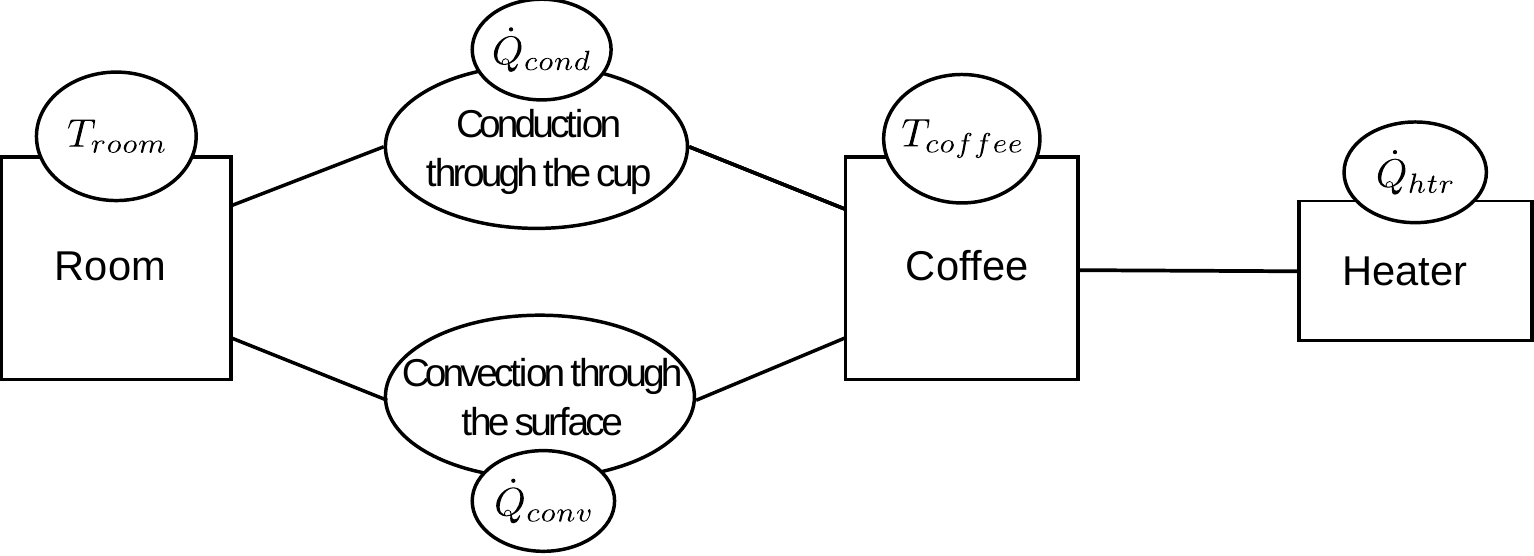}
 \caption{Case study 2: Modeling a cup of coffee in a room with a manual coffee heater.}
 \label{fig:example_coffee_room_heating_manual}
 \end{figure}

 We extend our first case study by adding a heater that is always on
 and therefore sends a constant flow of heat to the cup of coffee, as
 illustrated in Fig.~\ref{fig:example_coffee_room_heating_manual}.
 Furthermore, we start with an initial coffee temperature of
 $-10^\circ C$, so that this example shows how the coffee entity
 exhibits a \textit{hybrid} behavior, when given a sufficient
 constant heat, going through all three main phases: solid, liquid,
 and gas.

We now declare the @coffee@ object to be a @WaterEntity@ object, and add an object 
@boiler@  of  class \texttt{HeatGenerator}. The other objects (@room@, @crConduct@, and
@crConvect@) are defined as in Section~\ref{sec:case1}:

\myttsize
\begin{alltt}
op cs2\! :\! -> GlobalSystem .
eq cs2 =
  \texttt{\char123}< coffee\! :\! WaterEntity | heatCap\! :\! coffeeHC, mass\! :\! coffeeMass, temp\! :\! \emph{-10}, 
                           tempDisplay\! :\! "", heatFlow\! :\! 0, mode\! :\! default, phase\! :\! solid, 
                           heatTrans\! :\! 0, heatTransDisplay\! :\! "" >
   ...
   < boiler\! :\! HeatGeneration | entity\! :\! coffee, Qdot\! :\! 15/10 >\texttt{\char125} .
\end{alltt}
\normalsize

As for the first case study, we can simulate the above system and
plot the room and coffee temperatures as functions of time. 
Figure~\ref{fig:plotCS1-CS2}\subref{fig:plotCS2} shows the simulation 
up to time $1500$. We notice that  the coffee temperature remains constant 
during the transition phases (melting and evaporation), while it increases 
when the coffee is in liquid, solid,  or gaseous state. The fact that the melting 
temperature is somewhat greater than $0$ $^\circ C$ is a consequence 
of the approximation in the computation of the continuous behavior and the 
chosen \texttt{timeStep}: a bigger \texttt{timeStep} will lead to a coarser 
numerical approximation in the Euler algorithm. The room tends to become warmer, 
compared to the first case study; this is due to the presence of the heater, which, through 
the coffee, indirectly transfers heat to the room.

If we would like to know  how long  it takes for 
the iced coffee to start to melt,  we can use Real-Time Maude's 
\texttt{find earliest} command to find the first reachable state 
such that the coffee is  \texttt{melting}:

\myttsize
\begin{alltt}
Maude>\!\!\! \textit{(find\! earliest\! cs2 =>* \char123C:Configuration <\!\! coffee\!\! :\!\! WaterEntity\!\! |\!\! phase\!\! :\!\! melting\!\! >\char125\!\! .) }
\end{alltt}

\normalsize \noindent In $224ms$ Real-Time Maude returns a solution
showing that the iced coffee starts to melt after 22 seconds.

\subsection{Case Study 3: Keeping the Coffee Warm}  \label{sec:case3}

We keep the set-up of the second case study, but now we define
a more sophisticated heater, namely, one that senses the coffee
temperature at each time step and tries to keep the temperature of the
coffee between $70^\circ C$ and $80^\circ C$ by turning itself on and
off. We need to modify the previous model by:
\begin{itemize}
\item adding two instantaneous rules: one for turning the heater on when the coffee temperature is below
$70^\circ C$, and one for turning the heater off when the coffee is warmer than $80^\circ C$, and
\item adding a condition to the tick rule so that time does not advance when 
one of the above rules should be taken. 
\end{itemize}
For convenience, we define a subclass @SmartHeater@ of the class
@HeatGenerator@ for such smart heaters:

\myttsize
\begin{alltt}
class SmartHeater | status : OnOff, lowTemp : Rat, highTemp : Rat .
subclass SmartHeater < HeatGenerator .

sort OnOff .   ops on off : -> OnOff .
\end{alltt}
\normalsize

The rules for turning the heater on and off are immediate:

\myttsize
\begin{alltt}
crl [turnOff] : 
    < E\! :\! ThermalEntity | temp\! :\! T1 >
    < HG\! :\! SmartHeater | entity\! :\! E, status\! :\! \emph{on}, highTemp\! :\! T2 >
   =>
    < E\! :\! ThermalEntity |  >
    < HG\! :\! SmartHeater | status\! :\! \emph{off}, Qdot\! :\! 0 >
   if T1 >= T2 .

crl [turnOn] : 
    < E\! :\! ThermalEntity | temp\! :\! T1 >
    < HG\! :\! SmartHeater | entity\! :\! E, status\! :\! \emph{off}, lowTemp\! :\! T2 >
   =>
    < E\! :\! ThermalEntity |  >
    < HG\! :\! SmartHeater | status\! :\! \emph{on}, Qdot\! :\! 15/10 >
   if T1 <= T2 .
\end{alltt}
\normalsize

We also define a function @heatersOK@, which holds in a state when none of the 
above two rules can be applied,  and  modify the tick rule to take this into account:

\myttsize
\begin{alltt}
op heatersOK : Configuration -> Bool [frozen (1)] .

eq heatersOK(< HG\! :\! SmartHeater | entity\! :\! E, status\! :\! S, lowTemp\! :\! T1, highTemp\! :\! T2 >
             < E\! :\! ThermalEntity | temp\! :\! T >  REST) 
     = ((S == on and T < T2) or (S == off and T > T1)) 
       and heatersOK( < E\! :\! ThermalEntity | >  REST) .
eq heatersOK(CONFIG) = true [owise] .

crl [tick]\! :\! \char123CONFIG\char125 => \char123computeTemp(computeQdot(CONFIG), timeStep)\char125 in time timeStep 
            if timeCanAdvance(CONFIG) \emph{and heatersOK(CONFIG)} .
\end{alltt}
\normalsize

The initial state is as in the second case study, with the exception
that  the ``dumb''
heater has been replaced by a smart heater:

\myttsize
\begin{alltt}
op cs3\! :\! -> GlobalSystem .
eq cs3 =
  \texttt{\char123}< coffee\! :\! WaterEntity | temp\! :\! -20, phase\! :\! liquid, ... >
   < coffeeHeater\! :\! SmartHeater | entity\! :\! coffee, status\! :\! off, Qdot\! :\! 0, lowTemp\! :\! 70, 
                         \!\!         highTemp\! :\! 80 >
   ...\texttt{\char125} .
\end{alltt}
\normalsize 


\begin{figure}[t]
 \centering
 \includegraphics[width=0.5\textwidth]{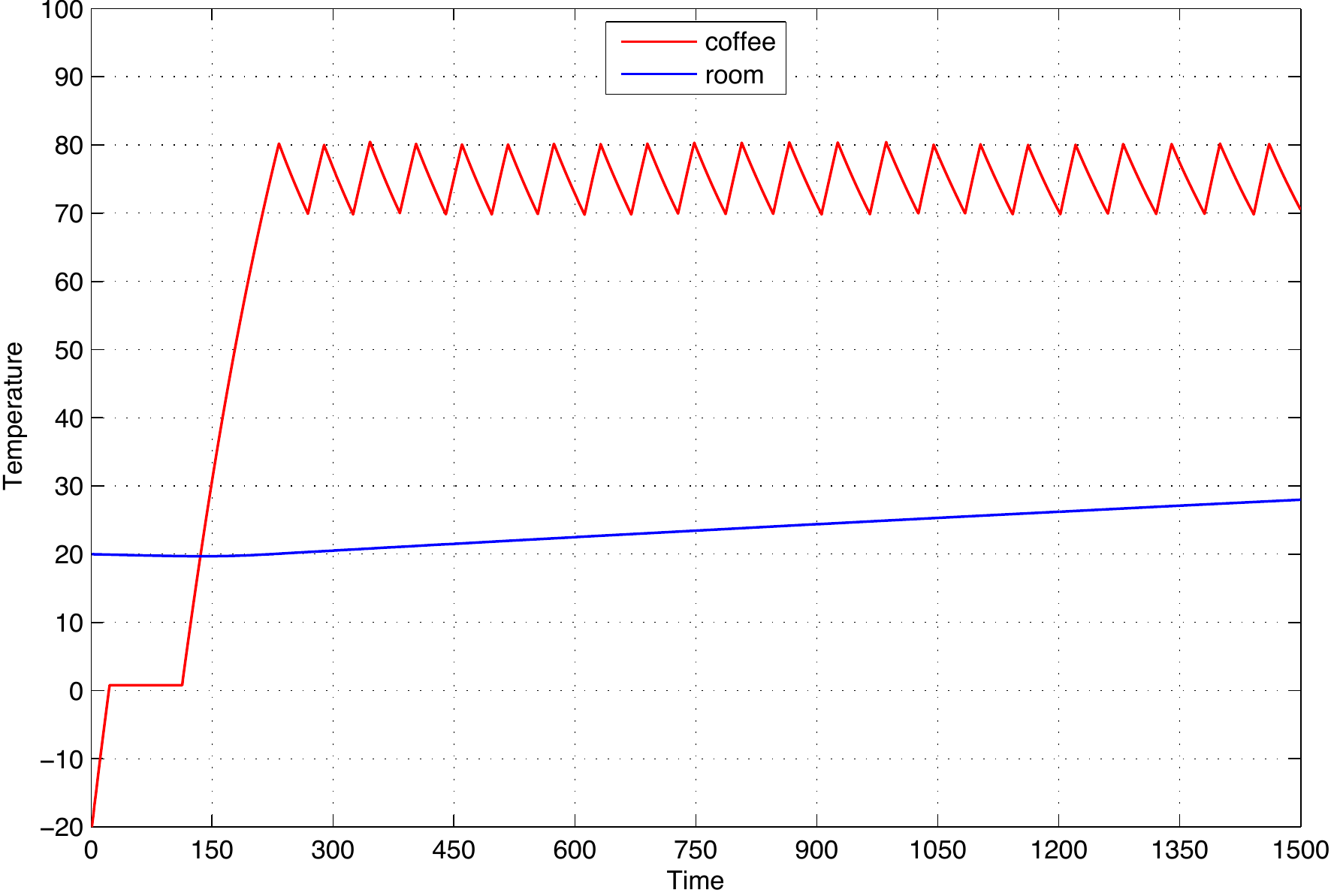}
 \caption{Case study 3: Simulation.}
 \label{fig:case3plot}
 \end{figure}

Figure~\ref{fig:case3plot} shows the coffee and room temperatures as
functions of the elapsed time 
in a Real-Time Maude simulation of this system. 

Finally, we use time-bounded LTL model checking to 
analyze the stability property that once the temperature of the coffee
has reached $69.5^\circ C$, its temperature will remain in the
interval  between 
$69.5^\circ C$  and $80.5^\circ C$ (we set the interval as 69.5 to
80.5 instead of 70 to 80 for taking into account the imprecision from
the approximation using the Euler method).  We  define the
atomic proposition \texttt{temp-ok}  to hold in all
states where the coffee temperature is between $69.5^\circ C$  and $80.5^\circ C$:

\myttsize
\begin{alltt}
op temp-ok : -> Prop [ctor] .
ceq \char123REST \;<\! coffee\! :\! WaterEntity\! |\! temp\! :\! T\! >\char125 |= temp-ok = (T >= 139/2 and T <= 161/2)\! .
\end{alltt}
\normalsize

The property can be checked by executing using time-bounded model checking:

\myttsize
\begin{alltt}
Maude> \emph{(mc cs3 |=t [] (temp-ok -> [] temp-ok) in time <= 1500 .)}

Result Bool :
  true
\end{alltt}
\normalsize

\section{Related Work}  \label{sec:related-work}

Modelica~\cite{fritzson} is an object-oriented language for modeling
physical systems where  hybrid
differential algebraic equations  model the continuous dynamics.
  The language supports the
specification of linear and non-linear equations.  Since the language
does not have a formal semantics, the precise meaning  of a model
depends on the simulation tool used. Furthermore, there are  no
reachability
and temporal logic analysis tools for Modelica. 

HyVisual~\cite{lee_zheng06} is an actor-based tool environment for the
simulation of continuous and hybrid systems.  Hybrid behaviors are
specified as finite state machines in which  a state can be refined into a
dynamic system represented as ordinary differential equations. 
HyVisual  does not provide 
model checking analyses. 

OO-DPT~\cite{villani} is an object-oriented  approach for
modeling hybrid supervisory systems.  This approach proposes a Petri-net-based formal
technique named \emph{object-oriented differential predicate
  transition net} (OO-DPT net).  This technique defines an interface
between differential equation systems and Petri nets to model hybrid
behaviors.  Reachability and safety properties analysis is the target
of this approach, but as of yet, there is no tool support
for OO-DPT. 

HyTech~\cite{henzinger_etal_97} is a tool for modeling and
verifying hybrid  systems represented as linear hybrid automata. The
discrete behaviors are represented as a set of locations and discrete
transitions between them where the continuous dynamics is defined in each
location.  The continuous dynamics is specified as polyhedral
differential inclusions.  The restrictions related to the formalism
used by the tool confines the representation of continuous dynamics to
the ones usually used in modeling physical systems. The main
difference between HyTech and Real-Time Maude is the expressiveness
and generality of the formalism, where in Real-Time Maude all kinds of
data types, functions, communication models, dynamic object creation,
etc., can be specified.

SHYMaude (Stochastic Hybrid
Maude)~\cite{meseguer_sharykin} is  a modeling language for object-based stochastic
hybrid systems that extends the  PMaude~\cite{kumar_etal}
 tool for probabilistic rewrite theories.  Stochastic
hybrid systems consist of distributed 
stochastic hybrid objects that interact with each other by
asynchronous message passing.  The continuous behaviors are governed
by stochastic differential equations, and the discrete changes are
probabilistic.  For simulation, the continuous dynamics of stochastic differential
equations is approximated by using the Euler-Maruyama method, a
generalization of Euler method to approximate numerical solution of
stochastic differential equations.  For formal analyses, the interface
of the statistical model checking tool VeStA~\cite{vesta} to Maude
system is used.
Despite the obvious similarities that both approaches are
object-oriented approaches to hybrid systems in extensions of Maude,
our work and the SHYMaude work are in fact quite different.
In~\cite{meseguer_sharykin}
the hybrid objects are autonomous and only interact with other
hybrid objects by message passing, whereas in our work, the main focus
is on  the \emph{physical interaction} (such as heat flow) between
the hybrid objects.

\section{Concluding Remarks}  \label{sec:concl}

We have presented a general object-based method for formally modeling
and analyzing thermal systems in Real-Time Maude.  In contrast to most other approaches, 
we also focus on the 
physical interactions between different physical components. We
explain how the Euler method can be adapted to approximate continuous
behaviors, and   have illustrated our method  on three variations of a
simple coffee-and-room system with realistic parameters. 

This work should be seen as a first exploration into how hybrid
systems can be modeled and analyzed in Real-Time Maude. Although we
believe that Real-Time Maude -- because of its expressiveness and
generality, support for objects, user-definable data types, different
communication models, etc. -- should be a suitable candidate for
modeling and analyzing advanced hybrid systems; this of course has to be
validated by applying our techniques on advanced
 hybrid systems. 

Euler is a simple numerical algorithm for
solving ordinary differential equations (ODE). It could be interesting
to implement in Real-Time Maude some other numerical approaches for
solving ODE (e.g. the midpoint method, the linear multi-step method,
or the Runge-Kutta method)  that offer more precise results than Euler,
probably at the cost of computational efficiency.  
Another point to explore is the possibility  of using the
other time sampling strategies offered by Real-Time Maude, like the
\textit{maximum time advance} strategy; applying this strategy to
continuous behavior would require an execution strategy that can
``predict'' the maximum time that the system can advance, before some
instantaneous rule has to be taken, possibly using an advanced
numerical method for ODE with \textit{dynamic step}.  

\paragraph{Acknowledgments.} We thank the anonymous reviewers for 
helpful comments on a previous version of this paper, and 
gratefully acknowledge financial support by the Research Council of Norway
through the Rhytm project, and by the Research Council of Norway and the 
German Academic Exchange Service (DAAD) through the DAADppp project 
 "Hybrid Systems Modeling and Analysis
with Rewriting Techniques (HySmart)."

\normalsize
\bibliographystyle{eptcs}
\bibliography{cm_bibliography,bibl}

\begin{thebibliography}{10}
\providecommand{\bibitemstart}[1]{\bibitem{#1}}
\providecommand{\bibitemend}{}
\providecommand{\bibliographystart}{}
\providecommand{\bibliographyend}{}
\providecommand{\url}[1]{\texttt{#1}}
\providecommand{\urlprefix}{Available at }
\providecommand{\bibinfo}[2]{#2}
\bibliographystart

\bibitemstart{bentley}
\bibinfo{author}{J.P. Bentley} (\bibinfo{year}{2005}):
  \emph{\bibinfo{title}{Principles of Measurement Systems}}.
\newblock \bibinfo{publisher}{Pearson Education Limited}, \bibinfo{edition}{4th
  ed.} edition.
\bibitemend

\bibitemstart{maude-book}
\bibinfo{author}{M.~Clavel}, \bibinfo{author}{F.~Dur\'an},
  \bibinfo{author}{S.~Eker}, \bibinfo{author}{P.~Lincoln},
  \bibinfo{author}{N.~Martí-Oliet}, \bibinfo{author}{J.~Meseguer} \&
  \bibinfo{author}{C.~Talcott} (\bibinfo{year}{2007}):
  \emph{\bibinfo{title}{All About {Maude} - A High-Performance Logical
  Framework}}, {\sl \bibinfo{series}{Lecture Notes in Computer Science}}
  \bibinfo{volume}{4350}.
\newblock \bibinfo{publisher}{Springer}.
\bibitemend

\bibitemstart{fritzson}
\bibinfo{author}{P.~Fritzson} (\bibinfo{year}{2003}):
  \emph{\bibinfo{title}{Principles of Object-Oriented Modeling and Simulation
  with Modelica}}.
\newblock \bibinfo{publisher}{Wiley-IEEE Computer Society Pr}.
\bibitemend

\bibitemstart{henzinger_etal_97}
\bibinfo{author}{T.A. Henzinger}, \bibinfo{author}{P.~Ho} \&
  \bibinfo{author}{H.~Wong-Toi} (\bibinfo{year}{1997}):
  \emph{\bibinfo{title}{HYTECH: A Model Checker for Hybrid Systems}}.
\newblock In: {\sl \bibinfo{booktitle}{CAV '97: Proceedings of the 9th
  International Conference on Computer Aided Verification}},
  \bibinfo{publisher}{Springer-Verlag}, \bibinfo{address}{London, UK}, pp.
  \bibinfo{pages}{460--463}.
\bibitemend

\bibitemstart{hoffman}
\bibinfo{author}{J.D. Hoffman} (\bibinfo{year}{2001}):
  \emph{\bibinfo{title}{Numerical Methods for Engineers and Scientists}}.
\newblock \bibinfo{publisher}{Marcel Dekker, Inc}, \bibinfo{edition}{2nd}
  edition.
\bibitemend

\bibitemstart{mike-wsn}
\bibinfo{author}{M.~Katelman}, \bibinfo{author}{J.~Meseguer} \&
  \bibinfo{author}{J.~Hou} (\bibinfo{year}{2008}):
  \emph{\bibinfo{title}{Redesign of the {LMST} Wireless Sensor Protocol through
  Formal Modeling and Statistical Model Checking}}.
\newblock In: \bibinfo{editor}{G.~Barthe} \& \bibinfo{editor}{F.~de~Boer},
  editors: {\sl \bibinfo{booktitle}{Formal Methods for Open Object-Based
  Distributed Systems (FMOODS'08)}}, {\sl \bibinfo{series}{Lecture Notes in
  Computer Science}} \bibinfo{volume}{5051}, \bibinfo{publisher}{Springer}, pp.
  \bibinfo{pages}{150--169}.
\bibitemend

\bibitemstart{kumar_etal}
\bibinfo{author}{N.~Kumar}, \bibinfo{author}{K.~Sen},
  \bibinfo{author}{J.~Meseguer} \& \bibinfo{author}{G.~Agha}
  (\bibinfo{year}{2003}): \emph{\bibinfo{title}{A Rewriting Based Model for
  Probabilistic Distributed Object Systems}}.
\newblock In: {\sl \bibinfo{booktitle}{Proc.\ Formal Methods for Open
  Object-based Distributed Systems (FMOODS'03)}}, {\sl \bibinfo{series}{Lecture
  Notes in Computer Science}} \bibinfo{volume}{2884},
  \bibinfo{publisher}{Springer}.
\bibitemend

\bibitemstart{lee_zheng06}
\bibinfo{author}{E.A. Lee} \& \bibinfo{author}{H.~Zheng}
  (\bibinfo{year}{2006}): \emph{\bibinfo{title}{HyVisual: A Hybrid System
  Modeling Framework Based on Ptolemy II}}.
\newblock In: {\sl \bibinfo{booktitle}{IFAC Conference on Analysis and Design
  of Hybrid Systems}}.
\newblock \urlprefix\url{http://chess.eecs.berkeley.edu/pubs/54.html}.
\bibitemend

\bibitemstart{norm-paper}
\bibinfo{author}{E.~Lien} \& \bibinfo{author}{P.~C. {\"O}lveczky}
  (\bibinfo{year}{2009}): \emph{\bibinfo{title}{Formal Modeling and Analysis of
  an IETF Multicast Protocol}}.
\newblock In: {\sl \bibinfo{booktitle}{Proc.\ Seventh IEEE International
  Conference on Software Engineering and Formal Methods (SEFM 2009)}},
  \bibinfo{publisher}{IEEE}.
\bibitemend

\bibitemstart{meseguer_sharykin}
\bibinfo{author}{J.~Meseguer} \& \bibinfo{author}{R.~Sharykin}
  (\bibinfo{year}{2006}): \emph{\bibinfo{title}{Specification and Analysis of
  Distributed Object-Based Stochastic Hybrid Systems}}.
\newblock In: {\sl \bibinfo{booktitle}{Proc.\ Hybrid Systems: Computation and
  Control (HSCC'06)}}, {\sl \bibinfo{series}{Lecture Notes in Computer
  Science}} \bibinfo{volume}{3927}, \bibinfo{publisher}{Springer}, pp.
  \bibinfo{pages}{460--475}.
\bibitemend

\bibitemstart{fase06}
\bibinfo{author}{P.~C. {\"O}lveczky} \& \bibinfo{author}{M.~Caccamo}
  (\bibinfo{year}{2006}): \emph{\bibinfo{title}{Formal Simulation and Analysis
  of the {CASH} Scheduling Algorithm in {R}eal-{T}ime {M}aude}}.
\newblock In: \bibinfo{editor}{L.~Baresi} \& \bibinfo{editor}{R.~Heckel},
  editors: {\sl \bibinfo{booktitle}{Fundamental Approaches to Software
  Engineering (FASE'06)}}, {\sl \bibinfo{series}{Lecture Notes in Computer
  Science}} \bibinfo{volume}{3922}, \bibinfo{publisher}{Springer}, pp.
  \bibinfo{pages}{357--372}.
\bibitemend

\bibitemstart{journ-rtm}
\bibinfo{author}{P.~C. {\"O}lveczky} \& \bibinfo{author}{J.~Meseguer}
  (\bibinfo{year}{2007}): \emph{\bibinfo{title}{Semantics and Pragmatics of
  {Real-Time Maude}}}.
\newblock {\sl \bibinfo{journal}{Higher-Order and Symbolic Computation}}
  \bibinfo{volume}{20}(\bibinfo{number}{1-2}), pp. \bibinfo{pages}{161--196}.
\bibitemend

\bibitemstart{aer-journ}
\bibinfo{author}{P.~C. {\"O}lveczky}, \bibinfo{author}{J.~Meseguer} \&
  \bibinfo{author}{C.~L. Talcott} (\bibinfo{year}{2006}):
  \emph{\bibinfo{title}{Specification and Analysis of the {AER/NCA} Active
  Network Protocol Suite in {R}eal-{T}ime {M}aude}}.
\newblock {\sl \bibinfo{journal}{Formal Methods in System Design}}
  \bibinfo{volume}{29}(\bibinfo{number}{3}), pp. \bibinfo{pages}{253--293}.
\bibitemend

\bibitemstart{ogdc-tcs}
\bibinfo{author}{P.~C. {\"O}lveczky} \& \bibinfo{author}{S.~Thorvaldsen}
  (\bibinfo{year}{2009}): \emph{\bibinfo{title}{Formal Modeling, Performance
  Estimation, and Model Checking of Wireless Sensor Network Algorithms in
  {R}eal-{T}ime {M}aude}}.
\newblock {\sl \bibinfo{journal}{Theoretical Computer Science}}
  \bibinfo{volume}{410}(\bibinfo{number}{2-3}), pp. \bibinfo{pages}{254--280}.
\bibitemend

\bibitemstart{vesta}
\bibinfo{author}{K.~Sen}, \bibinfo{author}{M.~Viswanathan} \&
  \bibinfo{author}{G.~Agha} (\bibinfo{year}{2005}): \emph{\bibinfo{title}{On
  Statistical Model Checking of Stochastic Systems}}.
\newblock In: {\sl \bibinfo{booktitle}{Proc.\ Computer Aided Verification
  (CAV'05)}}, {\sl \bibinfo{series}{Lecture Notes in Computer Science}}
  \bibinfo{volume}{3576}, \bibinfo{publisher}{Springer}, pp.
  \bibinfo{pages}{266--280}.
\bibitemend

\bibitemstart{villani}
\bibinfo{author}{E.~Villani}, \bibinfo{author}{P.~E. Miyagi} \&
  \bibinfo{author}{R.~Valette} (\bibinfo{year}{2007}):
  \emph{\bibinfo{title}{Modelling and Analysis of Hybrid Supervisory Systems: A
  Petri Net Approach}}.
\newblock \bibinfo{publisher}{Springer}.
\bibitemend

\bibitemstart{wellstead}
\bibinfo{author}{P.~E. Wellstead} (\bibinfo{year}{1979}):
  \emph{\bibinfo{title}{Introduction to physical system modelling}}.
\newblock \bibinfo{publisher}{London : Academic Press}.
\bibitemend

\bibliographyend
\end{thebibliography}
\normalsize

\end{document}